\documentclass[12pt]{article}

\usepackage{amssymb}

\begin{document}

\begin{titlepage}

\begin{flushright}
arXiv:0805.0781
\end{flushright}
\vskip 2.5cm

\begin{center}
{\Large \bf Limits on Neutron Lorentz Violation from the\\
Stability of Primary Cosmic Ray Protons}
\end{center}

\vspace{1ex}

\begin{center}
{\large Brett Altschul\footnote{{\tt baltschu@physics.sc.edu}}}

\vspace{5mm}
{\sl Department of Physics and Astronomy} \\
{\sl University of South Carolina} \\
{\sl Columbia, SC 29208 USA} \\

\end{center}

\vspace{2.5ex}

\medskip

\centerline {\bf Abstract}

\bigskip

Recent evidence appears to confirm that the ultra-high-energy primary cosmic ray
spectrum consists mostly of protons. The fact that these protons can traverse
large distances to reach Earth allows us to place bounds on Lorentz violations.
The protons neither emit vacuum Cerenkov radiation nor $\beta$-decay into neutrons,
and this constrains six previously unmeasured coefficients in the neutron
sector at the $5\times 10^{-14}$ level. 
Among the coefficients bounded here for the first time are those that control
spin-independent boost anisotropy for neutrons. 
This is a phenomenon which could have existed (in light of the preexisting bounds)
without additional fine tuning.
There are also similar bounds for others species of hadrons. The bounds on
Lorentz violation for neutral pions are particularly strong, at the
$4\times 10^{-21}$ level, eleven orders of magnitude better than previous
constraints.

\bigskip

\end{titlepage}

\newpage

The observation~\cite{ref-abbasi} of the Greisen-Zatsepin-Kuz'min
(GZK) cutoff~\cite{ref-greisen,ref-zatsepin} and
the discovery that ultra-high-energy cosmic rays are associated with
nearby active galactic nuclei~\cite{ref-abraham} resolved a major
puzzle in physics. A number of exotic physical scenarios had
been suggested to explain the apparent absence of the GZK cutoff in
earlier observations. Such exotic physics would now seem to be
unnecessary; however, it is interesting to turn things around and ask what
constraints
can be placed on exotic theories based on our improved understanding of
cosmic ray physics.

One exotic idea which has attracted significant interest in the last
decade (and which was put forward to explain the apparent anomalies in the
cosmic ray spectrum~\cite{ref-coleman2}) is Lorentz violation. There has already been
some discussion of how new ultra-high-energy cosmic ray data can be used to
constrain Lorentz violation~\cite{ref-gagnon,ref-carone,ref-klinkhamer2}. However,
there is a great deal
more useful information that can still be extracted from what we
now know about the highest-energy cosmic rays.

Lorentz violation can be described by an effective field theory, the
standard model extension (SME), which contains all possible local
Lorentz-violating operators constructed from standard model
fields~\cite{ref-kost1,ref-kost2}. The minimal SME, which is the standard
theory used to parameterize Lorentz tests, is a version of the SME
containing only gauge invariant and power-counting renormalizable
operators. Many of the minimal SME coupling constants have been
constrained extremely tightly, but others have not.
(For details, see~\cite{ref-reviews2}. Up-to-date information
on minimal SME bounds can be found in~\cite{ref-tables}.)
Here we shall show
that cosmic ray observations allow us to place constraints on several
previously unbounded parameters. Among these are six SME coefficients in
the neutron sector, including three which parameterize spin-independent
boost invariance violations (a phenomenon which has never before been studied in
this sector).

It may seem surprising that bounds on any neutron coefficients are possible,
since neutron physics ordinarily have little to do with primary cosmic
rays. However, Lorentz violation could allow otherwise forbidden proton-to-neutron
transitions. We therefore wish to emphasize the rather interesting observation that
measurements that are entirely concerned with one species of particles (here,
protons) can be used to place strong constraints on exotic physics involving
different particles (neutrons or other species).

The experimental confirmation that the GZK cutoff does indeed exist at
$\sim6\times10^{10}$ GeV is good evidence that ultra-high-energy primary
cosmic rays are mostly protons. That protons should dominate the spectrum
was expected originally, but was called into question by the observation
of a few cosmic ray events well above the cutoff. However, we can now feel
fairly confident (although not absolutely certain) that we understand the
protonic cosmic ray spectrum up to the highest energies.

We shall assume that most of the highest energy primary cosmic rays are indeed
protons. If
this somehow turns out not to be the case, the constraints derived here would be
invalidated. Obviously then, further confirmation of the protonic nature of the
rays would be helpful.

We will also encounter another caveat that constrains the applicability of
our bounds. Cosmic ray data imply that the Lorentz violation coefficients
from the minimal SME must
obey a large number of one-sided inequalities. In order to translate these into
two-sided bounds on the individual coefficients, an assumption must be made about
the form of the Lorentz violation. This assumption is that isotropic Lorentz
violation (pure boost noninvariance) is  sufficiently small. This assumption
is not presently justified by experiment, but it will nonetheless be made in
much of our analysis. Making the assumption allows us to see much more easily
which forms of Lorentz violation cosmic ray measurements are sensitive to. This will
be discussed in more detail below.

The minimal SME Lagrange density for each spin-$\frac{1}{2}$
fermion species includes terms of the form
\begin{equation}
\label{eq-L}
{\cal L}=\bar{\psi}[i(\gamma^{\mu}+c^{\nu\mu}\gamma_{\nu}-d^{\nu\mu}\gamma_{\nu}
\gamma_{5})\partial_{\mu}-m]\psi.
\end{equation}
There is a different set of $c$ and $d$ coefficients for each species.
There are many additional terms in the most general minimal SME ${\cal L}$, but $c$
and $d$ are the most relevant at high
energies.
The effects of other terms are suppressed in relative importance by powers of $m/E$.
There are also analogous coefficients for scalar and vector particles.
Since Lorentz violation is known to be a small effect, all the calculations here will
be done to leading order in the violation coefficients.

When Lorentz symmetry is broken by $c$ or $d$, novel effects may appear.
For example, it may become possible for sufficiently energetic protons to decay.
Obviously, this is a
unique signature of
Lorentz violation. In a Lorentz-symmetric theory, if the decay is forbidden
when the proton is at rest, it is forbidden for all proton momenta; and even if the
proton is not absolutely stable at rest, Lorentz invariance dictates that it will
survive longer when it is highly boosted, because of the time dilation effect.

Of interest are decays of the proton into two or more particles, $p^{+}
\rightarrow n+\pi^{+}$, for example. At high energies, the Lorentz violation
changes the energy-momentum relation for each particle involved to
\begin{equation}
E_{w}(\vec{p}\,)=\sqrt{m_{w}^{2}+[1+2\delta_{w}(\hat{p})]\vec{p}\,^{2}},
\end{equation}
where $w$ labels the species, and the maximum achievable velocity (MAV) for a
species in the direction $\hat{p}$ is $1+\delta_{w}(\hat{p})$. Other forms of
Lorentz violation may modify the energy-momentum relation at lower energies,
but any such effects are suppressed in relative importance by powers of
$m_{w}/p$. The parameter $\delta_{w}$ can be spin dependent, and for
spin-$\frac{1}{2}$ fermions, it is equal to~\cite{ref-altschul4}
\begin{equation}
\delta_{w}(\hat{p})=-c^{w}_{00}-c^{w}_{(0j)}\hat{p}_{j}-c^{w}_{jk}\hat{p}_{j}
\hat{p}_{k}+sd^{w}_{00}+sd^{w}_{(0j)}\hat{p}_{j}+sd^{w}_{jk}\hat{p}_{j}
\hat{p}_{k},
\end{equation}
where $s$ is the helicity, $c_{(0j)}=c_{0j}+c_{j0}$ [likewise for $d_{(0j)}$], and
the superscripts indicate that the coefficients are those for the species $w$.

Proton decay above a certain threshold could occur if there were a mismatch
between the $\delta_{p}$ for the proton and the corresponding parameters
for the daughter particles. We shall consider three possible kinematic configurations
for this kind of proton decay. In each configuration, all the Lorentz-violating
effects are governed by a single parameter $\delta_{p}(\hat{p})-\delta_{w}(\hat{p})$.
The first configuration isolates the $c$ and $d$
coefficients for a single decay product by having the other decay products be
nonrelativistic; this makes it possible to obtain clean
bounds on the coefficients for just that one species of particle. The second
configuration is relevant if it is known that there is Lorentz violation in one
sector only. This is an interesting case to consider theoretically, but the analysis
does not lead to rigorous bounds on any physical coefficients. The third kinematic
configuration applies when the proton does not actually decay, but rather emits a
light
neutral particle, as in vacuum Cerenkov radiation $p^{+}\rightarrow p^{+}+\gamma$, or
$p^{+}\rightarrow p^{+}+\pi^{0}$.

In the first configuration, one of the
daughter particles will carry away practically all of the proton's initial
momentum $\vec{p}$, leaving the remaining one (or more) essentially at rest.
This is generally not the way the momentum is divided up at the threshold for
the reaction, but this configuration ensures that only the $\delta_{w}$ for a single
species among the daughter particles will enter our calculations.
The moving particles are essentially collinear, so only the $\delta_{w}(\hat{p})$
values for a single direction enter.
Energy-momentum conservation requires that for a two-body decay at threshold,
\begin{equation}
\label{eq-Econs}
(1+\delta_{p})p=(1+\delta_{1})p+m_{2},
\end{equation}
where 1 and 2 label the daughter particles.
We have dropped terms that are suppressed relative to the particle masses by
powers of $m/p$. Moreover, because only one direction is involved, we have
omitted the dependence of $\delta_{w}$ on $\hat{p}$.

Apparently, this process can occur at a momentum $p$ if
$\delta_{p}-\delta_{1}=m_{2}/p$. A positive $\delta_{p}$ indicates that
the initial proton with momentum $p$ has more energy than it would have in the
Lorentz-invariant theory. If this momentum is transferred to a particle with
a smaller $\delta_{1}$, there may be enough energy left over to create a
particle of type 2. If $\delta_{p}<\delta_{1}$, however, the process will
never be allowed, so the observed absence of these kinds of decays can only
give one-sided bounds on the various $\delta_{p}-\delta_{w}$. However, one-sided
bounds on the $\delta_{w}(\hat{p})$ for different directions $\hat{p}$ can be used to
place two-sided bounds on some of the individual $c^{w}$ and $d^{w}$ coefficients.

The generalization to more than two decay products is straightforward. If one
particle still carries away all the momentum, the
process can occur if $\delta_{p}-\delta_{1}>(\sum m_{w})/p$, where
$\sum m_{w}$ is the sum of the masses of all the other daughter particles.
Usually this sum will be dominated by the mass of the heaviest product other than the
one carrying away the momentum. It follows that the strength of the bounds we
can place on the SME coefficients relevant to a particular species is determined
by the masses of the other species that are produced along with it. The bounds for a
particular species are better if the other particles involved are lighter.

Of particular interest is the three-body $\beta$-decay $p^{+}\rightarrow n+e^{+}
+\nu_{e}$. The fact that protons of energy $E$ traveling in the direction $\hat{p}$
do not decay in this fashion places a bound
\begin{equation}
\label{eq-bound}
\delta_{p}(\hat{p})-\delta_{n}(\hat{p})<\frac{m_{e}}{E},
\end{equation}
and this can be used to constrain some quantities in the neutron sector that have
not previously been measured. For the most energetic observed cosmic rays,
$m_{e}/E\lesssim 10^{-14}$. It is interesting that the constraint (\ref{eq-bound})
is only ${\cal O}(m/E)$, whereas most previous bounds on $\delta_{w}$ coming from
high-energy observations were ${\cal O}(m^{2}/E^{2})$; this new energy dependence
arises from the involvement of a nonrelativistic particle in the decay reaction.

The kinematic configuration in which all but one of the daughter particles is at
rest is not generally the threshold configuration. However, it is a configuration
in which the $\delta_{w}$ for only one decay products enters. If it is known that
there is Lorentz violation for only one of the daughter particles, then the true
threshold can be determined straightforwardly. While calculating this threshold, we
shall assume that the particle with the Lorentz-violating dispersion relation is
much heavier than the other particles produced alongside it, as is the
neutron in $p^{+}\rightarrow n+e^{+}+\nu_{e}$.

In the threshold configuration, the heavy, Lorentz-violating particle again
carries away most of the momentum, so it will be moving at an
ultrarelativistic speed. However, since the other daughter particles are
substantially less massive, they too can be highly relativistic, even though they
carry only a small fraction of the total momentum.
At threshold, the moving particles are again essentially collinear.

To lowest order, the lighter particles produced in the decay may actually be
neglected. Neglecting them, the threshold at
which the conversion of a proton into the heavy decay product becomes allowed is
given approximately by
\begin{eqnarray}
\sqrt{m_{p}^{2}+(1+2\delta_{p})p^{2}} & = & \sqrt{m_{1}^{2}+(1+2\delta_{1})p^{2}} \\
(1+\delta_{p})p+\frac{m_{p}^{2}}{2p} & = & (1+\delta_{1})p+\frac{m_{1}^{2}}{2p}.
\end{eqnarray}
We have dropped terms that are suppressed beyond lowest order in $m/p$.
The process can occur if
$\delta_{p}-\delta_{1}=(m_{1}^{2}-m_{p}^{2})/2p^{2}$, placing the threshold at
approximately $p_{T}=\sqrt{(m_{1}^{2}-m_{p}^{2})/2(\delta_{p}-\delta_{1})}$.

If the other decay products are light enough, they will have little impact on the
threshold value of $p$. If $m_{w}\ll\sqrt{m_{1}^{2}-m_{p}^{2}}$,
then including an additional particle of species $w$ among the decay products only
raises the threshold momentum by $\sqrt{2/(\delta_{p}-\delta_{1})}m_{w}$. The
additional
momentum is divided evenly between the heavy particle 1 and the light particle $w$.
Each light daughter particle increases the threshold only slightly, with the net
result that the threshold occurs at
\begin{equation}
p_{T}=\sqrt{\frac{2}{\delta_{p}-\delta_{1}}}\left(\frac{\sqrt{m_{1}^{2}-m_{p}^{2}}}
{2}+\sum m_{w}\right).
\end{equation}

Unfortunately,
this result is not very useful for placing bounds, because the analysis presupposes
that Lorentz violation is only important for the heaviest decay product. If the
values of $\delta_{w}$ for the other decay products are comparable to the
$\delta_{1}$ corresponding to the heaviest product, the overall scale of the
threshold is still $p_{T}\sim\sqrt{(m_{1}^{2}-m_{p}^{2})/(\delta_{p}-\delta_{1})}$.
However, if Lorentz violation for the other particles is allowed to be larger, the
threshold may increase dramatically. If the $\delta_{w}$ for the lighter particles
grows large enough, the threshold configuration may become the first one considered,
in which all but the heaviest particle are produced at rest.

If there were known to be no Lorentz violation in the electron or neutrino sectors,
the analysis of the true threshold location would give a much tighter bound on
$\delta_{p}-\delta_{n}$ than (\ref{eq-bound}). The bounds on $\delta_{p}-\delta_{n}$
would be at the $10^{-22}$ level.
However, Lorentz violation in the
neutrino sector is very poorly constrained. Therefore, (\ref{eq-bound})
represents the best bound that can be placed at the present time.

The third kinematic configuration we shall consider is one with a proton and a
lighter neutral particle in the final state. Because there are only two species
involved in the whole process, the threshold will again depend only on the combination $\delta_{p}-\delta_{1}$. When the particles are collinear, as they are
at threshold, energy-momentum conservation dictates
\begin{equation}
\label{eq-Epcons}
(1+\delta_{p})p+\frac{m_{p}^{2}}{2p}=(1+\delta_{p})(p-p_{1})+\frac{m_{p}^{2}}
{2(p-p_{1})}+(1+\delta_{1})p_{1}+\frac{m_{1}^{2}}{2p_{1}}.
\end{equation}
The threshold lies at the minimum of $p$ as a function of $p_{1}$; setting
$dp/dp_{1}=0$ determines $\delta_{p}-\delta_{1}$ to be
\begin{equation}
\delta_{p}-\delta_{1}=\frac{m_{p}^{2}}{2(p_{T}-p_{1})^{2}}-\frac{m_{1}^{2}}
{2p_{1}^{2}}.
\end{equation}
With this $\delta_{p}-\delta_{1}$, and $m_{1}/m_{p}\ll 1$, energy-momentum
conservation is satisfied for $p_{1}=(2m_{1}^{2}/m_{p}^{2})^{1/3}p_{T}\ll p_{T}$. At
threshold, the momentum carried away by the lighter particle is small compared with
the momentum that remains with the proton.

If protons up to momentum $p$ do not radiate neutral particles in this way, we must
therefore
have $\delta_{p}-\delta_{1}<m_{p}^{2}/2p^{2}$.
This bound has already been recognized
for vacuum Cerenkov radiation~\cite{ref-klinkhamer2}, for which case
$m_{1}=m_{\gamma}=0$. The Cerenkov threshold can also be calculated using the usual
result that Cerenkov radiation is
emitted when a charged particle moves faster than the phase speed of light in the
same direction. We shall discuss the vacuum Cerenkov bounds in more detail shortly.
However, the absence of the process $p^{+}\rightarrow p^{+}+\pi^{0}$ can also be used
to place bounds on Lorentz violation in the pion sector, which are significantly
better than any previous pion bounds. For heavier uncharged mesons, with masses
comparable to $m_{p}$, the above analysis must be modified slightly, but the resulting bounds are at a similar level---worse only by an ${\cal O}(1)$ factor
if there are no weak interactions involved in the process.

Having discussed the decay kinematics in several scenarios, we shall turn to
determining the quantitative bounds on the $\delta$ parameters for various particles.
Bounds on neutron parameters are of the greatest interest. However, there is one
remaining issue, related to how long the various decay or emission processes we have
discussed will take.
Obviously, for bounds be based on a given reaction process, it must be possible for
that reaction to occur
in the travel time of a cosmic ray proton. However, this criterion is easily
satisfied for any reaction that is mediated by either the strong or electromagnetic
interaction. The rate of vacuum Cerenkov radiation was calculated
in~\cite{ref-altschul9}, and the process is extremely rapid. A strong process
such as $p^{+}\rightarrow p^{+}+\pi^{0}$ will naturally occur even more quickly.

However, since the decay $p^{+}\rightarrow n+e^{+}+\nu_{e}$ is a weak interaction
process, there might be a concern that it will not be rapid enough to
occur
during the primary proton's flight to Earth. Yet a proton traveling
10 Mpc to Earth with an energy of $10^{11}$ GeV experiences a proper time
span
greater than $10^{4}$ s. This is much longer than the typical lifetime of
weakly decaying hadrons, longer in particular than the neutron lifetime of
approximately 880 s.

In a Lorentz invariant theory,
the rate at which a process occurs is determined by the matrix element, which
depends on the details of the dynamics, and the phase space available to the
outgoing particles. The rate is generally a rapidly increasing function of the
available phase space. These statements continue to hold in theories with Lorentz
violation~\cite{ref-kost5}. The rate at which a proton decays is determined by the
matrix element for the decay and the phase space available to the decay products in
the initial proton's rest frame. In this frame, the Lorentz violation actually makes
only a small correction to the phase space---provided that the energy available for
the decay is known. The phase space $\Pi^{w}$ for a given decay product is equal to
$\Pi^{w}_{0}[1+{\cal O}(\delta_{w})]$, where $\Pi^{w}_{0}$ is the phase space for
a Lorentz invariant particle with the same energy. For determining the approximate
rate for a process, it is therefore a reasonable approximation to ignore the
effects of Lorentz violation on the decay rate, except in the calculation of how
much energy is released in the decay.

The amount of energy that is available for the daughter particles to carry away as
kinetic energy is quite straightforward to calculate. To determine this available
energy, we can (as in the second kinematical configuration considered above)
begin by neglecting all particles produced in the decay except the heaviest one.
In the laboratory frame, where the original proton is moving extremely rapidly,
the heavy decay product's energy is $E_{w}=(1+\delta_{w})p+\frac{m_{w}^{2}}{2p}$. The
energy of this configuration in the proton's rest frame is $E_{r}=\gamma_{p}(E_{w}-
v_{p}p)$, where $v_{p}$ and $\gamma_{p}$ are the proton's velocity and Lorentz
factor in the laboratory frame. To the required order, $v_{p}=(1+\delta_{p})-\frac
{m_{p}^{2}}{2p^{2}}$ and $\gamma_{p}$ is simply $p/m_{p}$. In terms of the
threshold momentum $p_{T}$ (which is the momentum at which $E_{r}=m_{p}$),
\begin{equation}
E_{r}=\frac{m_{p}^{2}-m_{w}^{2}}{2m_{p}}\frac{p^{2}}{p_{T}^{2}}+
\frac{m_{p}^{2}+m_{w}^{2}}{2m_{p}}.
\end{equation}
When this is less than the rest energy $m_{p}$ of the proton, the decay products will
carry some kinetic energy in the center of mass frame.
The amount of kinetic energy available to the decay products in this frame is
\begin{equation}
\label{eq-availE}
m_{p}-E_{r}=\frac{m_{2}^{2}-m_{p}^{2}}{2m_{p}}\left(\frac{p^{2}}{p_{T}^{2}}-1\right)
\approx(m_{w}-m_{p})\left(\frac{p^{2}}{p_{T}^{2}}-1\right),
\end{equation}
where the approximation in (\ref{eq-availE}) is valid if $m_{w}\approx m_{p}$, which
holds if the particle $w$ is a neutron. The inclusion of the other, lighter decay
products decreases the available energy by a small amount.

If $p=2p_{T}$, the available energy is greater than the neutron-proton mass
difference of 1.29 MeV. Therefore, the phase space available for the decay at this
energy is greater than the phase space for the decay of a stationary neutron.
The matrix element for the proton $\beta$-decay process $p^{+}\rightarrow n+e^{+}+
\nu_{e}$ near threshold is extremely similar to the usual matrix element
for neutron $\beta$-decay. The matrix element is not suppressed by the smallness of
the Lorentz violation.
In fact, if the available energies for the processes are
precisely the same (that is, if $m_{p}-E_{r}=m_{n}-m_{p}$), the only differences
between the rates for the two reactions
would come from small changes to the phase space and isospin-violating differences
in the matrix elements. Since the available phase space and rate for the
proton disintegration increase with $p$, above $p=2p_{T}$, the proton will decay
with a mean proper
lifetime less than 880 s. With a lifetime this short, nearly all the
primary protons with more than twice the threshold energy will decay during their
journeys. The weak nature of the decay is no impediment to the decay's ready
occurrence during the time it takes for a cosmic ray proton to reach Earth.

We can therefore use (\ref{eq-bound}) to place rigorous new bounds on Lorentz
violation for neutrons. However, in order to establish new bounds, we
must understand the existing bounds in both the neutron and proton sectors.
We shall first consider the existing bounds based on vacuum Cerenkov radiation.
If this process were allowed above some
threshold, a proton with a higher energy would radiate away the excess energy
extremely quickly~\cite{ref-altschul9}.
If protons with energies up to $E$ are observed not to emit vacuum Cerenkov
radiation, this translates into a bound on the quantity $\delta_{p}-\delta_{\gamma}$
that is ${\cal O}(m_{p}^{2}/E^{2})$. The quantity
$\delta_{\gamma}(\hat{p})$
is analogous to the fermionic $\delta(\hat{p})$, and it
governs the phase speed of photons in the direction $\hat{p}$.
$\delta_{\gamma}(\hat{p})$
depends on the $k_{F}$ coefficients in the Lorentz-violating Lagrange density for
the electromagnetic sector,
\begin{equation}
{\cal L}_{A}=-\frac{1}{4}F^{\mu\nu}F_{\mu\nu}-\frac{1}{4}k_{F}^{\mu\nu\rho\sigma}
F_{\mu\nu}F_{\rho\sigma}.
\end{equation}
However, the spin-dependent part of $\delta_{\gamma}$ has been shown to be extremely
small using measurements of cosmological
birefringence~\cite{ref-kost11,ref-kost21,ref-kost22},
and it may therefore be neglected. Moreover, for any one sector, the
spin-independent part of $\delta$ may be defined away. We shall use this freedom
to make the electromagnetic sector conventional.
Our results can be adapted to a different convention with a nonzero $k_{F}$ simply by
making the replacement
$c^{\mu\nu}\rightarrow c^{\mu\nu}-\frac{1}{2}k_{F\alpha}\,^{\mu\alpha\nu}$ in
every matter sector.

Existing cosmic ray data place bounds on the components of
$\delta_{p}-\delta_{\gamma}$, which is just $\delta_{p}$ with our conventions. If
protons of both helicities are part of the
primary cosmic ray spectrum, these bounds apply to both helicities; if only one
helicity is present, the bounds obviously apply only to that helicity. What is
important is that any proton that is part of the primary cosmic ray flux is
subject to these bounds.

Most of
the details of the vacuum Cerenkov bounds are worked out in~\cite{ref-klinkhamer2}.
From the absence of vacuum Cerenkov radiation,
there are two-sided bounds on the $-c^{p}_{(0j)}+sd^{p}_{(0j)}$ and $-c^{p}_{jk}
+sd^{p}_{jk}$
coefficients appearing in $\delta_{p}$---if it is assumed that the isotropic
term $-c^{p}_{00}+sd^{p}_{00}$ is small compared with the others.
Two-sided bounds in the presence of a generic $c^{p}_{00}$ are impossible, because a
large enough
$c^{p}_{00}$ can render the proton MAV less than one for every direction and spin.
The absence of vacuum Cerenkov radiation thus cannot constrain a positive
$c^{p}_{00}$.
However, if $-c^{p}_{00}+sd^{p}_{00}\approx0$, the other coefficients are all
separately
bounded at the $2\times10^{-21}$ level.
(If the anisotropic terms are assumed to vanish instead,
there is a one-sided bound on the isotropic coefficient,
$c^{p}_{00}-sd^{p}_{00}>-10^{-22}$.)

There are also bounds on $c^{p}$ and $d^{p}$ (as well
as neutron coefficients) from laboratory experiments with atomic clocks---some of which
are much stronger than the cosmic ray bounds, ranging from the $10^{-20}$ to $10^{-29}$
levels~\cite{ref-kost6,ref-bear,ref-cane,ref-wolf,ref-kornack}---although most of the
$d^{p}$ coefficients remain unconstrained.
Yet laboratory experiments which rely on the
Earth's rotation to search for anisotropic effects are typically insensitive to a number
of SME coefficients.
Moreover, the clock comparison
experiments using hyperfine transitions have only been sensitive to forms of
neutron boost invariance
violations that also depend on the
spin. The $c^{n}_{(Tj)}$ coefficients, which characterize a spin-independent anisotropy in the
way neutrons respond to Lorentz boosts, have never been studied.

If the isotropic part of the proton Lorentz violation is small, the preexisting bounds
justify setting
$\delta_{p}=0$ in (\ref{eq-bound}), because the vacuum Cerenkov bounds in the proton
sector are quite a bit tighter than the new bounds we shall be considering here.
The remaining neutron bounds coming from (\ref{eq-bound}), which have the form
$-\delta_{n}(\hat{p})<
m_{e}/E$, are based on precisely the same data set as the vacuum Cerenkov bounds on
$\delta_{p}$. So the consequences for the individual neutron $c$ and $d$ coefficients
are (unsurprisingly) similar. If there is no isotropic neutron term, each of the
neutron
coefficients $-c^{n}_{(0j)}+sd^{n}_{(0j)}$ or $-c^{n}_{jk}+sd^{n}_{jk}$ is bounded
above and below by $5\times10^{-14}$. On the other hand, if there is exclusively
isotropic neutron Lorentz violation (but still no isotropic proton Lorentz
violation), $c^{n}_{00}-sd^{n}_{00}<10^{-14}$. What is more, these bounds
must hold for both helicities: $s=\pm1$. This did not need to be the case for
the proton bounds, since only one proton helicity had to be immune to vacuum Cerenkov
radiation for some protons to reach Earth; however, if the
proton $\beta$-decay process
were allowed for even one daughter neutron helicity, the primary protons would all
decay away. Invoking the bounds for both signs of $s$, we see that actually each
$|c^{n}_{(0j)}|+|d^{n}_{(0j)}|$ or $|c^{n}_{jk}|+|d^{n}_{jk}|$ is bounded
by $5\times 10^{-14}$.

This gives us our primary result for the neutron sector.
If the dominant forms of Lorentz violation are not isotropic, then the previously
unmeasured neutron coefficients $\frac{1}{4}|c^{n}_{Q}|=\frac{1}{4}|(c^{n}_{XX}+
c^{n}_{YY}-2c^{n}_{ZZ})|$, $|c^{n}_{(TX)}|$, $|c^{n}_{(TY)}|$, $|c^{n}_{(TZ)}|$,
$\frac{1}{2}|d^{n}_{(XZ)}|$, and $|d^{n}_{(TZ)}|$ are all bounded by
$5\times10^{-14}$. $X$, $Y$, $Z$,
and $T$ are the coordinates in the sun-centered celestial equatorial
coordinate system in which bounds on Lorentz violation are conventionally
expressed. If isotropic
Lorentz violation is possible, then the optimal bounds are represented by the
one-sided inequalities (\ref{eq-bound}), with the
$\hat{p}$ and $E$ values for each ultra-high-energy proton cosmic ray that has been
observed. There is still sensitivity to (just not two-sided bounds on) the six
neutron coefficients just mentioned.

Of course, the bounds are not
actually limited to the neutron sector. Any baryon that can be
produced via the $\beta$-decay of a proton is subject to similar bounds. This means
baryons with charge 0 or +2. There are bounds for non-baryonic charged particles as
well, but they are weaker. The reason for this is that there must be a baryon among
the daughter particles, and if this is not the particle of interest, its mass will
dominate the $\sum m_{w}$ on which the strength of the bound depends. Because
baryons are so much heavier than the electron, the bounds
are worse by a factor of $2\times 10^{3}$ for charge $+1$ particles (which
can be created in reactions that have nucleons in the final states) or slightly more
for charge $-1$ and $-2$ particles (which must be produced along with something
heavier, like a $\Delta$).
The resulting bounds on the $\delta_{w}$ parameters for various charged mesons and
baryons are at the $10^{-10}$ level. This makes them slightly worse than bounds
derived from certain other astrophysical observations for charged
pions~\cite{ref-altschul16};
however, they are an improvement over previous bounds for almost all heavier species
of hadrons~\cite{ref-altschul14}. (The
$\Delta^{+}$ is an exception; there are much better bounds on the MAV for
this species because of its involvement with the GZK cutoff.)

Neutral particles that do not carry baryon number and can be produced singly by the
strong interaction are a separate case. The absence of the process
$p^{+}\rightarrow p^{+}+\pi^{0}$ gives strong new bounds on the Lorentz violation
coefficients for neutral pions. The relevant Lagrangian for the pion field is
\begin{equation}
{\cal L}_{\pi}=\frac{1}{2}(\partial^{\mu}\pi)(\partial_{\mu}\pi)+\frac{1}{2}
k^{\pi}_{\mu\nu}(\partial^{\mu}\pi)(\partial^{\nu}\pi)
-\frac{m^{2}}{2}\pi^{2}.
\end{equation}
The coefficients $k^{\pi}$ are analogous to the fermionic $c$, and
$\delta_{\pi}(\hat{p})=-\frac{1}{2}\left[k^{\pi}_{00}+2k^{\pi}_{j0}\hat{p}_{j}+
k^{\pi}_{jk}\hat{p}_{j}\hat{p}_{k}\right]$.
Again assuming the isotropic component of the Lorentz violation is negligible, there
are two-sided bounds on all the remaining $k^{\pi}$ coefficients, which must be
smaller than $4\times 10^{-21}$. These represent an eleven order of magnitude
improvement over the neutral pion bounds from~\cite{ref-altschul16}.
There should be similar (but slightly weaker) bounds for any heavier neutrals mesons
that can be produced strongly. These include the pseudoscalar $\eta$ and $\eta'$, as
well as the vector $\rho^{0}$, $\omega$, and $\phi$. However, the full set of
Lorentz-violating coefficients relevant for a massive, spin-1 particle has not
been worked out.

If primary cosmic ray protons were undergoing $\beta$-decays, the positrons produced
in the decay could also become part of the cosmic ray spectrum. The fact that there
are observed to be very few electrons and positrons among the most energetic cosmic
rays might be used to place bounds on the Lorentz violation coefficients for
the electron sector. However, the resulting bounds would be at
only the $10^{-10}$ level,
significantly weaker than other astrophysical bounds on the same
coefficients~\cite{ref-altschul6,ref-altschul15}, which constrain $c^{e}$ and $d^{e}$
at the $10^{-14}$--$10^{-17}$ levels.

This whole analysis presumes that it is known that the cosmic rays we observe are
largely protons. However, the arguments that the spectrum is predominantly
protonic are usually made in the context of a theory that is assumed to be Lorentz
invariant. If primary cosmic ray protons above a certain energy may decay into
neutrons, which are stable at that energy, the spectrum might change from
proton-dominated to neutron-dominated above the energy in question. Put another way,
the decay $p^{+}\rightarrow n+e^{+}+\nu_{e}$ still has an energetic nucleon among
the decay products; above the threshold for this decay, the neutron that is
produced could simply replace the proton as part of the spectrum. While this is to
some extent an accurate evaluation, it is important to realize that the proton
decay would still leave a distinctive imprint on the observed cosmic ray spectrum.
The reason is that the decay does not simply transfer all the primary proton's
energy to the daughter neutron. The proton energy is split up between three
particles, and so the neutrons produced via the decay will possess less energy than
their protonic forbears. Moreover, if the initial proton is not too far above the
threshold for the decay, the neutron's energy may even be below the threshold; in
that case, the neutron itself will decay, back into an even lower energy proton. The
net process in that case is
$p^{+}\rightarrow p^{+}+e^{-}+e^{+}+\nu_{e}+\bar{\nu}_{e}$; the proton releases
some of its energy in the form of lepton-antilepton pairs. In any case, if above
some energy, the cosmic ray spectrum changed from protonic to neutronic, there would
be a pronounced feature in the spectrum at that energy. Since there are no large
unexplained features in the ultra-high-energy cosmic ray spectrum, we can safely
rule out this possibility.

There is also the unlikely possibility that the highest energy cosmic rays are not
protons, but rather are predominantly nuclei, most likely $^{56}$Fe. If this somehow
turns out to be the case, the bounds on neutron Lorentz violation discussed here
would be weakened somewhat. Proton to neutron conversion could still occur inside
the nuclei if $\delta_{p}-\delta_{n}$ were large enough,
but the resulting bounds would be less stringent,
primarily because of the smaller momenta
of the constituent hadrons. Each proton in an $^{56}$Fe cosmic ray with energy $E$
carries a momentum of roughly $E/56$. This decrease in momentum, in conjunction with
the effects of nuclear binding,
will probably lead to bounds that are worse
by about two orders of magnitude. However, the $^{56}$Fe scenario seems unlikely.
Heavier primary cosmic rays will generally produce air showers with more muons,
and this can be used to constrain the composition of the primary cosmic ray
spectrum~\cite{ref-risse}. Estimates of the fraction of $^{56}$Fe
nuclei among the highest energy cosmic rays vary, but~\cite{ref-dova} placed an upper
limit of 64\% on the iron fraction. It is also possible that the cosmic rays might be
produced initially as nuclei, which then undergo photodisintegration in flight.
However, this would not significantly effect our bounds on Lorentz violation.
If photodisintegration of $^{56}$Fe is the source of primary cosmic ray protons,
the protons will still propagate over megaparsec distances~\cite{ref-anchordoqui2}.
Since the mean lifetime
of protons with energies more than twice the decay threshold is significantly shorter
than intergalactic travel times, the protons would still have time to decay in large
numbers.

In summary, we have shown that the stability of primary cosmic ray protons with
energies up the GZK cutoff has implications for Lorentz violation. That there were
bounds on SME parameters coming from the protons' stability against vacuum Cerenkov
radiation had
already been observed. However, the fact that an energetic proton does not decay
into a neutron places new bounds on Lorentz violation coefficients in the neutron
sector, at the $m_{e}/E$ level. These bounds are most conveniently expressed in
the absence of isotropic Lorentz violation, in which case six previously unmeasured
neutron coefficients are bounded by $5\times 10^{-14}$; this includes the first
constraints on spin-independent forms of boost anisotropy for neutrons.
The absence of similar decay
processes place comparable bounds on the coefficients for other charged 
baryons, while there are weaker bounds for charged
hadrons that do not have baryon number $B=1$. For neutral mesons that can be
produced via the strong interaction, the bounds are significantly better---at the
$4\times10^{-21}$ level for neutral pions, for example.
All this demonstrates the
continued utility of high-energy astrophysical data for constraining Lorentz
violation and possibly other forms of exotic physics.


\end{document}